\documentclass[prl,aps,twocolumn,showpacs]{revtex4}
\usepackage{amsmath,amssymb,amsfonts,float}
\usepackage{hyperref}
\usepackage[dvips]{graphicx,color}    
\usepackage{times,pifont}       
\newcommand{\bolK}{\text{\bf K}}

\newcommand{\bolk}{\mathbf{k}}
\newcommand{\bolp}{\text{\bf p}}
\newcommand{\bolq}{\mathbf{q}}

\newcommand{\bolQ}{\mathbf{Q}}

\newcommand{\bolr}{\mathbf{r}}

\newcommand{\bra}[1]{\langle #1 |}  
\newcommand{\ket}[1]{| #1 \rangle}  
\newcommand{\VEV}[1]{\langle #1 \rangle}

\newsavebox{\dotdot}
\savebox{\dotdot}[3mm]{\shortstack{\circle*{0.8}\\ \\ \circle*{0.8}}}

\begin{document}
\title{Stability and magnetization curve of spin-nematic phase slightly below saturation field
}
\author{Hiroaki~T.~Ueda$^{1}$, and Keisuke~Totsuka$^2$}
\affiliation{
$^1$Okinawa Institute of Science and Technology, Onna-son, Okinawa 904-0412, Japan\\
$^2$Yukawa Institute for Theoretical Physics, Kyoto University, Kitashirakawa Oiwake-Cho, Kyoto 606-8502, Japan}
\begin{abstract}
We discuss the magnetization process slightly below the saturation field in 
frustrated magnets. A condensation of bound magnons on the spin-polarized state  
induces either a spin nematic phase or a state with phase separation. 
The (effective) interaction between the bound magnon pairs not only is crucial 
to the stability of the nematic phase, but also determines the slope of 
the magnetization curve near saturation. 
We generally derive the expression of this interaction by using the perturbative 
scattering theory. By applying the method to coupled zigzag chains LiCuVO$_4$, we find the positive 
pair-pair interaction implying the stability of the spin nematic phase.  
We also point out that the magnetization curve of 
LiCuVO$_4$ is almost vertical ({\em i.e.} very large $dM/dH$) near the saturation 
exhibiting one-dimensional feature despite non-negligible interchain couplings. 
\end{abstract}
\maketitle
{\it Introduction}-
Magnetic frustration sometimes brings about various exotic phenomena in quantum magnets. 
One such example  
is the appearance of the spin nematic phase, in which long-range magnetic order manifests itself not in 
the local magnetization but in the rank-2 tensor \cite{Andreev-G}.   
Although the possibility of spin nematic phases have been pointed out theoretically 
in various systems \cite{review_nem}, 
there are only a handful experimental candidates.  
A spin-1/2 compound LiCuVO$_4$, which may be viewed as coupled quantum
$S = 1/2$ $J_1$-$J_2$ chains \cite{LiCuVO4_1,LiCuVO4_2}, 
is one of the promising candidates that are supposed to realize  
the putative nematic phase under high magnetic fields $H$ \cite{NemExp_3,NemExp_4}.  
Close to saturation, stable bound states of spin-flip excitations (magnons) are formed 
by the ferromagnetic nearest-neighbor interaction 
and the spin-nematic phases realize when they condense earlier than the single magnon \cite{Chubukov}. 
Recently, slightly below (3.9-4.9T) the saturation field $H_{\text{sat}}= 44$T (52T) 
for $H$ parallel to the $c$ ($a$ or $b$) axis, an additional phase transition has been 
reported \cite{NemExp_3}; on the basis of the trial wave function, 
the new phase above it has been argued \cite{Zhitomirsk-Tsunetsugu} 
to be the spin nematic phase. 
As magnons are bosons, this phase may be viewed as a magnetic counterpart of the pair-superfluid 
phase of bosons \cite{Nozieres-S-82,Kuklov-P-S-04} and the coupled $J_1$-$J_2$ system provides us with 
a unique playground to study the physics of bosonic composite.

Theoretically, the single-chain model has been extensively investigated both numerically 
and analytically \cite{Hikihara,Sudan-L-L-08,Meisner}, 
and the results obtained have shed some lights on the behavior of the coupled-chain system 
LiCuVO$_4$.  
For instance, the spin modulated phase found experimentally 
\cite{NemExp_1,NemExp_2,Schrettle-08,Masuda-11,NemExp_4} 
may be viewed as the SDW2 phase (a bosonic density-wave phase formed by pairs) found in the DMRG studies.\cite{Hikihara}
On the other hand, the existence of a long-range helical spin order 
at $H=0$ implies non-negligible effects of interchain couplings. 
Recently, several attempts have been made \cite{HTUandKT,S.Nishimoto,Zhitomirsk-Tsunetsugu,MSato} 
to investigate the effects of the interchain couplings on the multipolar phases found 
in 1D \cite{Hikihara,Sudan-L-L-08}.  
One of the few, but reliable methods to study the nematic phases independent of the dimensionality is 
to calculate the energy of a bound state of magnons on 
the spin-polarized background \cite{Chubukov,Kecke,Zhitomirsk-Tsunetsugu,HTUandKT}. 

In this letter, we study the properties of the nematic 
phase formed slightly below the saturation field 
by developing a method of 
calculating the effective pair-pair interaction on the basis of the dilute Bose-gas expansion 
(see, {\em e.g.}, Ref.\cite{DiluteBose} for a review). 
The interaction among condensed bound magnons plays an important role; 
not only being crucial to the stability of the nematic phase, it also determines 
the asymptotic form of the magnetization curve as well as the critical temperature 
of the nematic phase. 
First we derive, for a general setting, an integral equation which determines this interaction up to first order 
in two-magnon-scattering amplitude.  
Next, we apply this method to the magnetic properties of a quasi-one-dimensional compound LiCuVO$_4$ 
under high magnetic field.  

\begin{figure}[H]
\begin{center}
\includegraphics[scale=0.25]{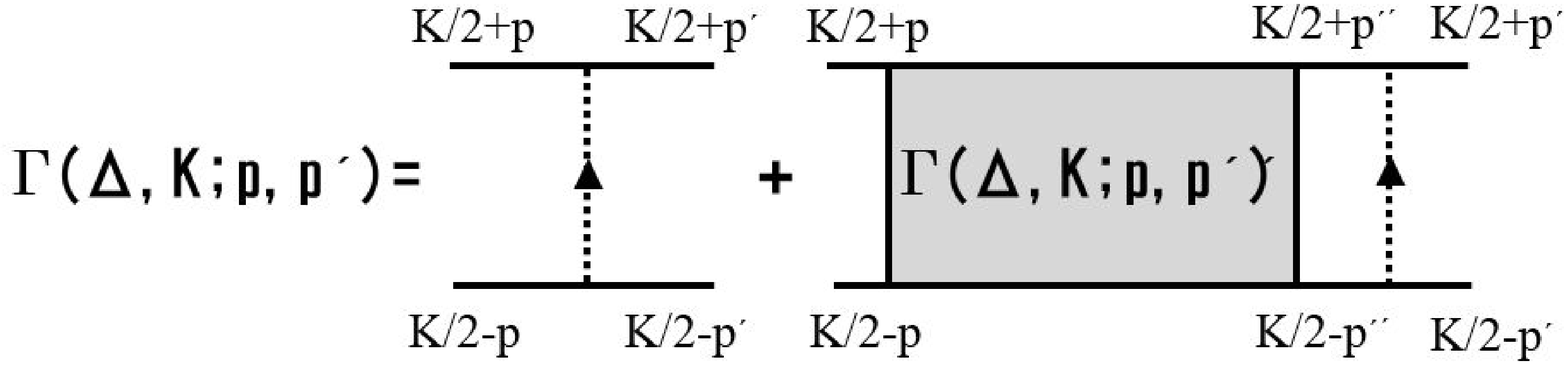}
\caption{\label{fig:scattering}
Magnon-magnon scattering amplitude $\Gamma$ given by the ladder diagram.
\label{Fig:scatteringM}}
\end{center}
\end{figure}

{\it Magnon Bose-Einstein Condensation (BEC)}-
Let us consider a spin-1/2 Heisenberg model on a Bravais lattice 
with generic interactions $\{ J_{ij} \}$ placed in a magnetic field ($H$):
\begin{equation}
\mathcal{H}=\sum_{\langle i,j\rangle}J_{ij}\,{\bf S}_i{\cdot}{\bf S}_j 
+ H \sum_{j}S^{z}_{j}\ .
\label{Ch1:SpinHamiltonian}
\end{equation}
We rewrite this Hamiltonian by the hardcore boson ($\uparrow$) representation of 
spin operator:
\begin{equation}
S^z_l=-1/2+a^\dagger_la_l \; , \;\;
S_l^+ =a_l^\dagger \; , \; \;  S_l^- =a_l
\label{Ch2:hardcoreSpin}
\end{equation}
to obtain the following boson Hamiltonian
\begin{equation}
\mathcal{H} = \sum_{q}(\omega (\bolq) - \mu)a^\dagger_{\bolq}a_{\bolq}
+\frac{1}{2N}  \sum_{\bolq,\bolk,\bolk^\prime}  V(\bolq) 
a_{\bolk+\bolq}^\dagger a_{\bolk^\prime-\bolq}^\dagger
a_{\bolk}a_{\bolk^\prime}\ ,
\label{Ch2:Hboson}
\end{equation}
\begin{equation}
\begin{split}
&\epsilon(\bolq)=\sum_{j} \frac{1}{2} J_{ij}\cos \left(
\bolq {\cdot}(\bolr_i-\bolr_j)\right)\ ,\ 
\omega(\bolq) =\epsilon(\bolq)-\epsilon_{\text{min}}\ ,\\
&\mu= H_{\text{c}1} -H \ ,\ 
H_{\text{c}1} =\epsilon({\bf 0})-\epsilon_{\text{min}}\ ,\ 
V(\bolq) =2(\epsilon(\bolq)+U)\ ,
\end{split}
\end{equation}
where $\epsilon_{\text{min}}$ is the minimum of the single-magnon energy $\epsilon({\bf q})$ 
and $U({\rightarrow}\infty)$ is the hard-core potential which guarantees $S=1/2$ at each site. 
Now, the external field $H$ controls the energy of a magnon as the chemical potential. 
If we reduce the magnetic field down to $H_{\text{c}1}$, the gap of a magnon closes 
and the single-magnon BEC may occur, 
which leads to $\VEV{S_l^{-}}=\VEV{a_l}\neq 0$ and thereby stabilizes various kinds of magnetic 
orders \cite{HTUandKT,Batyev}. 
Then, the emergent phase is determined \cite{HTUandKT} by the effective 
interaction between condensed magnons, 
which is given by the magnon-magnon scattering amplitude $\Gamma$ at 
$\Delta=0$
(see Fig.~\ref{Fig:scatteringM}):
\begin{equation}
\begin{split}
&\Gamma(\Delta,\bolK;\bolp,\bolp^\prime)=V(\bolp^\prime-\bolp)+V(-\bolp^\prime-\bolp)\\
&-\frac{1}{2}\int \frac{d^d p^{\prime\prime}}{(2\pi)^d}
\frac{\Gamma (\Delta,\bolK;\bolp,\bolp^{\prime\prime})
\left\{
V(\bolp^\prime-\bolp^{\prime\prime})+V(-\bolp^\prime-\bolp^{\prime\prime})
\right\}}{\omega(\bolK/2+\bolp^{\prime\prime})
+\omega(\bolK/2-\bolp^{\prime\prime})+\Delta -i0^+}\ ,
\end{split}
\label{Ch2:laddereq}
\end{equation}
where $\bolK$ and $\Delta$ respectively are 
the center-of-mass momentum of the two magnons in question and  
the binding energy. 
This integral equation is exactly solvable\cite{Batyev}.

{\it Magnon-pair condensation}-
If a stable bound state of magnons (a magnon pair) exists, the single-magnon BEC 
is not necessarily the leading instability from the spin-polarized state. 
In fact, if the gap of the magnon pair is smaller than double of that of the single-magnon, 
magnon-pair condensation occurs at $H_{\text{c2}}(> H_{\text{c1}})$ and the spin-nematic order 
$\VEV{S^\pm_l}=0,\ \VEV{S^+_lS^+_l}\neq0$ takes place. 
The binding energy of the bound state $\Delta_{\text{B}}(\bolK)$ 
is determined by a pole of the scattering amplitude $\Gamma$. 
The wave function of the bound state $\chi_\bolK(\bolp)$ 
then is obtained as the residue of $\Gamma$ at the pole \cite{Nakanishi-Bethe}.   

If we assume that the condensation of pairs occurs only at $\bolK=\bolQ_{\text{B}}$, 
the effective potential for the pair-superfluid phase may be written as 
\begin{equation}
\frac{1}{N}E(\rho_2) =\frac{1}{4}
\Gamma^{(2)}\rho_2^2-(\Delta_{\text{min}}+2\mu)\rho_2\ ,
\label{EA}
\end{equation}
where $\rho_2$ and $\Delta_\text{min}$ respectively are 
the density of the condensed pairs and 
the binding energy at the bottom of the pair dispersion 
$\Delta_\text{min}=\Delta_{\text{B}}(\bolQ_{\text{B}})$. 
The normalization constant $1/4$ in front of $\Gamma^{(2)}$ is introduced 
for the symmetry factor of the scattering amplitude.  
The interaction between the condensed pairs $\Gamma^{(2)}$  
is the only parameter that remains to be determined in this effective potential. 

The sign of $\Gamma^{(2)}$ determines the stability of the spin nematic 
phase: 
negative $\Gamma^{(2)}$ in eq.(\ref{EA}) naively means 
that $\rho_2 \rightarrow \infty$ is favored on the energetic ground 
if the higher order terms in $\rho_2$ are neglected. 
Then, real-space collapse of magnon pairs destroys the long-range nematic order and leads 
to the first-order transition at some $\mu_{\text{c}}$ satisfying $\Delta_{\text{min}}+2\mu_{\text{c}}<0$.  
On the other hand, if $\Gamma^{(2)}$ is positive, 
the dilute condensate of pairs is stable and 
the second-order phase transition at $\mu=-\Delta_{\text{min}}/2$ may occur. 
When $\mu>-\Delta_{\text{min}}/2$, by minimizing the potential 
and neglecting the non-condensate contribution, 
we obtain the asymptotic form of magnetization near $H=H_{\text{c2}}$: 
\begin{equation}
\VEV{S^z_l}+\frac{1}{2}=2\rho_2=\frac{4(\Delta_{\text{min}}+2\mu)}{\Gamma^{(2)}}
=\frac{8}{\Gamma^{(2)}}(H_{\text{c}2}-H)\ ,
\label{dMdH2}
\end{equation}
where $H_{\text{c}2}=H_{c1}+\Delta_{\text{min}}/2$ is the actual saturation field.
If we introduce the creation operator of the bound state  
$d_\bolK^\dagger=\sum_{\bolp}\chi_\bolK(\bolp)a^\dagger_{\bolK/2+\bolp}a^\dagger_{\bolK/2-\bolp}$, 
the low-energy dynamics near $\bolK\approx \bolQ_{\text{B}}$ 
may be described by the following effective Hamiltonian:
\begin{equation}
\begin{split}
H_{\text{eff}} = &\sum_{\bolK\sim\bolQ_{\text{B}}}
\left\{
\sum_{i=x,y,z}\frac{(K_i-Q_{\text{B} i})^2}{2m_i^{(2)}}-
\mu_2
\right\}
d^\dagger_{\bolK}d_{\bolK}\\
&+\frac{\Gamma^{(2)}}{4N}\sum_{\bolK_1,\bolK_2,\bolq}
d^\dagger_{\bolK_1+\bolq}d^\dagger_{\bolK_2-\bolq}d_{\bolK_1}d_{\bolK_2} 
+ \cdots \ ,
\end{split}
\label{eq:bmH}
\end{equation}
where $\mu_2=2(H_{\text{c}2}-H)$ and 
the ellipsis denotes higher-order interactions which are 
suppressed in the dilute limit. 
The mass $m^{(2)}_{x,y,z}$ of the bound state is 
obtained by expanding the pair dispersion $\Delta_{\text{B}}(\bolK)$ around its minimum. 
Then, in the dilute limit, the phonon spectrum in the condensed phase is given by
\begin{equation}
\Omega_2(\bolK)= \sqrt{\mu_2\sum_{i=x,y,z}
\frac{(K_i-Q_{\text{B}i})^2}
{m^{(2)}_{i}}}\ .
\label{Ch2:BMdisp}
\end{equation}
The dilute-Bose gas approximation is justified when 
$\Gamma^{(2)}(m_{x}^{(2)}m_{y}^{(2)}m_{z}^{(2)}\rho_{2})^{1/3} \ll1$.   


Now we are at the place to derive $\Gamma^{(2)}$ from the scattering process 
of the magnon pairs.  
In terms of $\Gamma^{(2)}$, the scattering process of two pairs at $\bolK=\bolQ_{\text{B}}$ 
(shown in the left panel of Fig.~\ref{Fig:boundInt}) may be expressed as
$(\frac{i}{E-\mu_2})^4 (-i\Gamma^{(2)})$,  
where $E\rightarrow\mu_2$ is assumed. 
\begin{figure}[ht]
\begin{center}
\includegraphics[scale=0.6]{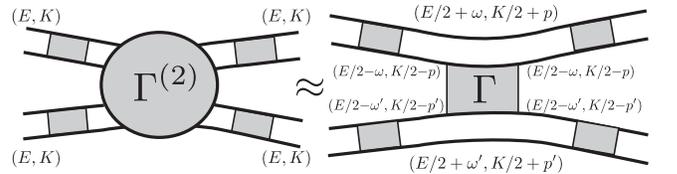}
\caption{Diagram of the two-body scattering of the magnon pairs. 
The hatched rectangles represent the two-body scattering amplitude $\Gamma$ of the single magnons 
(see Fig.\ref{Fig:scatteringM}). 
\label{Fig:boundInt}}
\end{center}
\end{figure}
By keeping only the first-order processes in $\Gamma$,  
we obtain the scattering amplitude $\Gamma^{(2)}$ as (see the right part of Fig.~\ref{Fig:boundInt})\cite{supple}
\begin{equation}
\begin{split}
&\Gamma^{(2)}=\int\frac{d^d\bolp d^d\bolp^\prime}{(2\pi)^{2d}}|\chi(\bolp)|^2|\chi(\bolp^\prime)|^2\\
\times &\Gamma
\bigl(
2\Delta(\bolQ_{\text{B}})+\omega(\bolQ_{\text{B}}/2+\bolp)+\omega(\bolQ_{\text{B}}/2+\bolp^\prime),\\
&\ \ \ \ \bolQ_{\text{B}}-\bolp-\bolp^\prime;(\bolp^\prime-\bolp)/2,(\bolp^\prime-\bolp)/2
\bigr)\ ,
\label{Ch2:effectiveIntBM}
\end{split}
\end{equation}
where the integration is carried out over the Brillouin zone 
and $\chi$ is normalized as $\int\frac{d^d\bolp}{(2\pi)^{d}}|\chi(\bolp)|^2=2$ 
since the bound state is the same for $\bolp$ and $-\bolp$. 
Our perturbation expansion in $\Gamma$ may be valid when $\Gamma$ (and hence, $\Gamma^{(2)}$, 
too) is small. 
The largeness of $\Delta_{\text{min}}$ 
may also be a simple criterion 
for the validity since $\Delta_{\text{min}}$ suppresses $\Gamma$ 
in eq. (\ref{Ch2:effectiveIntBM}). 

\begin{figure}[ht]
\begin{center}
\includegraphics[scale=0.3]{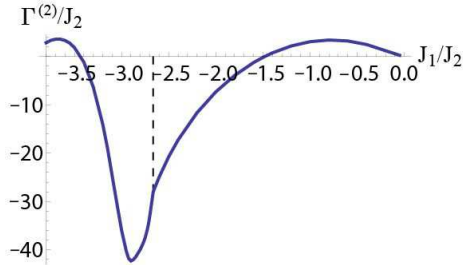}
\caption{(color online) Strength of magnon-pair interaction $\Gamma^{(2)}$ for the 1D $J_1/J_2$ chain.  
The kink occurs at  $J_1/J_2 = -2.67$ where the bottom of the bound state $Q_{\text{B}}$ 
changes from incommensurate value to commensurate one $Q_{\text{B}}=\pi$.  
\label{Fig:1DJ1J2_int}}
\end{center}
\end{figure}

{\it Application to 1D chain}-
First, let us apply our formulation to the $J_1$-$J_2$ spin chain, which is well studied by the DMRG 
and the bosonization technique \cite{Hikihara,Meisner}. 
The Hamiltonian is given by
\begin{equation}
\begin{split}
\mathcal{H}
=\sum_{i}\left(
J_1{\bf S}_{i}\cdot {\bf S}_{i+1}+J_2 {\bf S}_{i}\cdot{\bf S}_{i+2}
\right)
+ H \sum_{i}S_{i}^z\ ,
\end{split}
\end{equation}
where $J_1<0$, $J_2>0$. The result of the numerical calculation for $\Gamma^{(2)}$ 
is shown in Fig.~\ref{Fig:1DJ1J2_int}. 
For $-1.5 \lesssim J_1/J_2 < 0$, $\Gamma^{(2)}$ 
is positive and the nematic phase is stable,  
while for $-3.5\lesssim J_1/J_2 \lesssim -1.5$, 
$\Gamma^{(2)}$ is negative and the nematic phase is unstable. 
One possible consequence of the negative $\Gamma^{(2)}$ in 1D or 2D is 
the formation of further bound states of the pairs (e.g., quartets) due to infra-red fluctuations and 
their condensation. 
This can be explicitly seen by considering, as in eq.~(\ref{eq:M2R}), the ladder diagram 
using the Hamiltonian (\ref{eq:bmH}) 
\footnote{If $\Gamma^{(2)}$ is positive, 
the sign of the interaction does not change by the infra-red fluctuation.
Hence, the stability of the nematic phase itself can be judged 
from $\Gamma^{(2)}$ even in low-dimensional systems. }.

The previous results \cite{Kecke,Hikihara} 
suggest the nematic phase for $-2.7\lesssim J_1/J_2<0 $ and the condensation of 
the three-magnon bound states for $J_1/J_2\lesssim -2.7$. 
Hence, $\Gamma^{(2)}$ should remain positive for $-2.7\lesssim J_1/J_2$. 
This discrepancy may be due to our perturbative expansion. 
Although our method underestimates $\Gamma^{(2)}$ in 1D, 
it may be more reliable in 2D and 3D 
where quantum fluctuation is believed to be weaker than that in 1D  
and perturbative techniques may work better. 

\begin{figure}[ht]
\begin{center}
\includegraphics[scale=0.25]{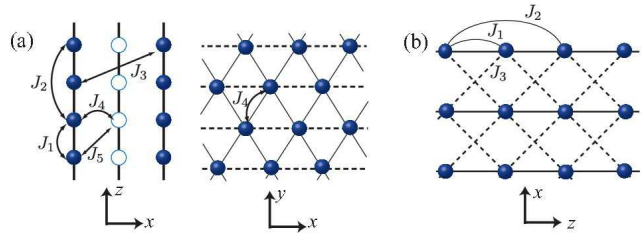}
\caption{(color online) (a) The lattice structure of LiCuVO$_4$  
and various exchange interactions \cite{LiCuVO4_2}. 
The circles denote spins ($S=1/2$). 
The dominant $J_1$-$J_2$ chains (thick lines) are running along the $z$($b$)-direction. 
(b) Neglecting relatively weak $J_4$ and $J_5$, one obtains 2D network (in $xz$-plane) 
considered here. 
\label{Fig:LiCuVO4}}
\end{center}
\end{figure}

{\it Application to LiCuVO$_4$}-
Next, we apply our method to LiCuVO$_4$ and,
by calculating $\Gamma^{(2)}$, determine magnetization curve 
near the saturation field . 
The lattice structure of LiCuVO$_4$ is shown in Fig.~\ref{Fig:LiCuVO4}
(the three crystal axes are $(x,y,z)\leftrightarrow (a,c,b)$). 
To avoid a long computational time, we neglected the weak 
$J_4$ and $J_5$ couplings and studied the following 
two-dimensional Hamiltonian in the $xz$-plane (see  Fig.~\ref{Fig:LiCuVO4}.(b)):
\begin{equation}
\begin{split}
&\mathcal{H}
= \sum_{\bolr}\bigl(
J_1{\bf S}_{\bolr} \cdot {\bf S}_{\bolr+\hat{\mathbf{e}}_z}
+J_2 {\bf S}_{\bolr} \cdot {\bf S}_{\bolr+2\hat{\mathbf{e}}_z} \\
& + J_{3}{\bf S}_{i}\cdot{\bf S}_{i+\hat{\mathbf{e}}_x + \hat{\mathbf{e}}_z}
+  J_{3}{\bf S}_{i}\cdot{\bf S}_{i+\hat{\mathbf{e}}_x - \hat{\mathbf{e}}_z}
\bigr)
+H \sum_{i} S_{i}^z\ ,
\end{split}
\label{2D:LiCuVO4}
\end{equation}
where $J_1=-1.6$ (meV), $J_2=3.8$ (meV) and $J_3=-0.4$ (meV) \cite{LiCuVO4_2} 
and $\hat{\mathbf{e}}_{x,z}$ denote the unit vectors in the direction of the crystal axes. 
For the values of $J_1,\ J_2,\ J_3$ given above, 
the minimal energy of the magnon pair $\Delta_{\text{min}} \approx0.12$ 
occurs at $\bolQ_{\text{B}}=(\pi,\pi)$ \cite{Zhitomirsk-Tsunetsugu}. 

\begin{figure}[ht]
\begin{center}
\includegraphics[scale=0.23]{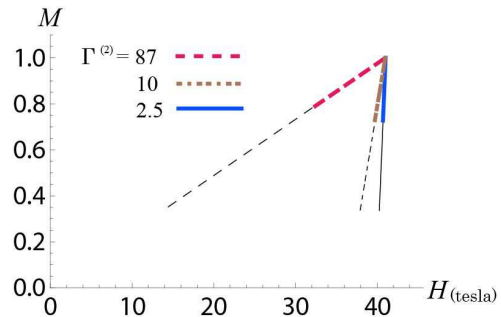}
\caption{(color online) Magnetization curves near saturation 
for various $\Gamma^{(2)}$.  
The thin lines are to underline the slopes.  
$\Gamma^{(2)}=2.5$ is the first-order result obtained for \eqref{2D:LiCuVO4}.  
$\Gamma^{(2)}=10$ is the upper bound of the renormalized $\Gamma^{(2)}$ (see Fig.~\ref{Fig:M2ladder}).  
The linear interpolation between the origin and saturation gives fairly large value $\Gamma^{(2)}\approx87$.  
\label{Fig:Magcurve}}
\end{center}
\end{figure}

The method described above yielded a positive value 
$\Gamma^{(2)}=2.5$ implying the stability of 
the spin-nematic phase near the saturation field.  
We plot the magnetization ($M=-2\VEV{S^z}$) curve obtained by eq.\eqref{dMdH2} in Fig.~\ref{Fig:Magcurve} 
(for $g(c\text{-axis})=2.3$ \cite{NemExp_3}, $H_{\text{c}2}\approx 41$ T).  
Clearly, due to small $\Gamma^{(2)}$, magnetization exhibits rapid decrease slightly below the saturation field, 
which is reminiscent of the pure-1D results\cite{Hikihara,Meisner}. 

{\it Effect of infra-red divergence}-
So far, we have studied specific models in 1D and 2D. 
However, physics of BEC strongly depends on the dimensionality. 
For example, if we could treat 
the scattering process up to infinite order  
as we do in the ladder approximation, 
infra-red fluctuations would suppress the scattering amplitude 
as $\Gamma^{(2)}_{\text{R}}\rightarrow 0$ for 1D and 2D models \cite{Fisher}. 
Hence, in low-dimensional systems, 
the magnetization curve 
has an infinite slope at the saturation field \cite{Sachdev-S-S-94}. 
In fact, from the renormalization group argument, 
the asymptotic form of magnetization 
is obtained as \cite{Fisher}  
\begin{equation}
\VEV{S_z}+\frac{1}{2}\propto 
\begin{cases}
(H_{\text{c}2}-H)\log|H_{\text{c}2}-H|\ ,\ & \text{for 2D},\\
(H_{\text{c}2}-H)^{\frac{1}{2}}\ ,\ & \text{for 1D} \; .
\end{cases}
\end{equation}
This implies that the effect of interplane (interchain) coupling in 2D case 
is more relevant than in 1D case. 
Actually, 
a small interplane coupling of the order of $J_p$ gives that
$dM/dH|_{H\approx H_c-0^+}\approx O(\log|J_p/J_{\text{intraplane}}|)+O(J_p^0)$: 
the steep behavior (log divergence) is fairly flattened by a week $J_p$.

\begin{figure}[ht]
\begin{center}
\includegraphics[scale=0.28]{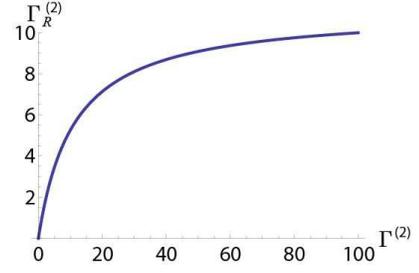}
\caption{(color online) Strong renormalization of pair-pair interaction $\Gamma^{(2)}$ 
by the infra-red fluctuations. 
We have assumed that the mass of the bound state 
is given by eq.~(\ref{mass_bound}) and the cutoff is given as 
$|K_{x,y,z}^{\prime}|<\Lambda=0.2$. 
\label{Fig:M2ladder}}
\end{center}
\end{figure}

To take into account the effects of infra-red fluctuations in our calculation,  
we consider the `super ladder diagram' constructed out of the process shown in the right part of 
Fig.~\ref{Fig:boundInt}. 
Now, we use the low-energy effective pair Hamiltonian eq.\eqref{eq:bmH}, 
which is tailored to LiCuVO$_4$, 
to calculate the two-magnon propagator,  
where we approximate $\Gamma^{(2)}$ by that obtained for the 2D model 
eq.~(\ref{2D:LiCuVO4})
\footnote{This approximation is valid 
when the binding energy $\Delta$ is enough larger
than the interplane coupling since $\Delta$ also suppresses 
the infra-red fluctuation}
and use the mass parameters of the bound state given by
\begin{equation}
m_x^{(2)}=110\ ,\ m_y^{(2)}=610000\ ,\ m_z^{(2)}=0.020\ .
\label{mass_bound}
\end{equation}
For comparison, we give those of a {\em single} magnon:  
$m^{(1)}_x=3.7$, $m^{(1)}_y=6.7$, and $m^{(1)}_z=0.068$. 
Surprisingly, we found that %
$m^{(2)}_z$ is by far smaller than $m^{(2)}_x$ and $m^{(2)}_y$  
implying that the low-energy dynamics is dominated by one-dimensional motion along 
the $z$-axis (i.e. $b$-axis) \footnote{In a quasi-low dimensional system, 
the dilute-Bose limit $\Gamma^{(2)}(m_{x}^{(2)}m_{y}^{(2)}m_{z}^{(2)}\rho_{2})^{1/3} \ll1$ 
is easily broken for a finite density of condensates $\rho_2$ due to a large mass. 
However, even in this case, $dM/dH_{H\rightarrow H_{\text{c}2}-0^{+}}$ is given by eq.~(\ref{dMdH2}), and we naively expect $dM/dH$ at a finite $\rho_2$ is analytically connected and 
continuous to the region of extremely-low $\rho_2$:
the deviation from our prediction using $M^{(2)}$ may be perturbative 
in $\rho_2$ unless an additional phase transition from the nematic phase occurs.}.
If we introduce the momentum cutoff $|K^\prime_{i}|\equiv|K_{i}-Q_{B,i}|<\Lambda$ $(i=x,y,z)$ into the 
Hamiltonian eq.~(\ref{eq:bmH}), 
the ladder diagrams for the pair-pair scattering are summed up to yield 
\begin{equation}
\Gamma^{(2)}_{\text{R}}=\frac{\Gamma^{(2)}}{1+\frac{\Gamma^{(2)}}{2}\int \frac{dK^{\prime 3}}{(2\pi)^3}
(\sum_{i}\frac{{K^\prime}_i^2}{2m_i})^{-1}}\ .
\label{eq:M2R}
\end{equation}
If we assume, for example, $\Gamma^{(2)}=2.5$ and the cutoff $\Lambda=0.2$,  
we obtain $\Gamma^{(2)}_{\text{R}}\approx 2.0$.  
The infra-red divergence suppresses the interaction 
as we expect from the results for the usual magnon-magnon scattering. 
This suppression is stronger for larger $\Gamma^{(2)}$ as is seen from Fig.~\ref{Fig:M2ladder}. 
In fact, we found that however large the bare pair-pair interaction $\Gamma^{(2)}$ 
might be, 
the renormalized value $\Gamma^{(2)}_{\text{R}}$ is bounded by a finite value about 10  
due to infra-red fluctuations.  
To summarize, the magnetization curve (near saturation) in general 
gets steeper when the fluctuation effects are included. 

{\it Comment on the collinear phase}-
Experimentally, it is believed that the high-field ($H\geq 8$T) phase  
with modulating $\VEV{S^z_l}$ and zero transverse magnetization \cite{NemExp_2,Schrettle-08,Masuda-11} 
is a 3D analogue of the SDW2 (a bosonic density-wave state of pairs) phase found in the 1D chain \cite{Hikihara,Meisner}.  
In 1D systems, either the spin-nematic phase or SDW2 is selected depending on the value of the effective 
Luttinger-liquid parameter \cite{Hikihara}.  
Below, we show that in higher dimensions it is possible to have a phase where the nematic order and modulating $\VEV{S^z_l}$ 
coexist. 

Before discussing the modulated phase, 
let us recall the nature of the nematic phase considered in the previous sections. 
The wave function of the pair condensate proposed 
in Ref. \cite{Zhitomirsk-Tsunetsugu} is:
$
|\text {BM}_1 \rangle = C_1 \exp (\rho_{\text{B}}\sum_{\bolp} 
\chi_{\bolK_1}(\bolp) a^\dagger_{\bolK_1/2+\bolp}a^\dagger_{\bolK_1/2-\bolp}) 
|\Omega\rangle
$ with 
$C_1 =\Pi_\bolp\sqrt{1-|\rho_{\text{B}}\chi_{\bolK_1}(\bolp)|^2}$ being the normalization constant.  
We can explicitly evaluate $\VEV{S^z_\bolr}$  
by using {\em e.g.}, $\VEV{a_\bolr}=0$ and 
$
\VEV{a^\dagger_\bolr a_\bolr}
=\frac{1}{N}\sum_{\bolp}\frac{|\rho_{\text{B}}
\chi_{\bolK_1}(\bolp)|^2}{1-|\rho_{\text{B}}\chi_{\bolK_1}(\bolp)|^2}=\text{const.}
$ 
to obtain a uniform magnetization. 
Therefore, the single-$\mathbf{K}$ condensation of the pairs leads to 
$\VEV{S^z_\bolr}=$const. 

On the other hand, if two modes (at $\bolK_1$ and $\bolK_2$) of bound states simultaneously 
condense (double-$\bolK$ condensation), the wave function may be given instead by
\begin{equation}
\begin{split}
|\text {BM}_2 \rangle &= C_2\exp \Bigl(\rho_{\text{B}1}\sum_{\bolp_1} 
\chi_{\bolK_1}(\bolp) a^\dagger_{\bolK_1/2+\bolp_1}a^\dagger_{\bolK_1/2-\bolp_1}\\
&+\rho_{\text{B}2}\sum_{\bolp_2} 
\chi_{\bolK_2}(\bolp) a^\dagger_{\bolK_2/2+\bolp_2}
a^\dagger_{\bolK_2/2-\bolp_2} \Bigr)|\Omega\rangle\ ,
\end{split}
\end{equation}
where 
$C_2=
\Pi_{\bolp_1,\bolp_2}\sqrt{1-|\rho_{\text{B}1}\chi_{\bolK_1}(\bolp_1)|^2}
$$\sqrt{1-|\rho_{\text{B}2}\chi_{\bolK_2}(\bolp_2)|^2}$ 
(with $\rho_{\text{B}i}$ the density of the $i$-th bound state). 
Then, the spin density is calculated as
\begin{equation}
\begin{split}
&\VEV{a^\dagger_\bolr a_\bolr}
\approx  |\rho_{\text{B}1}|^2+|\rho_{\text{B}2}|^2
+\biggl\{
\exp (i(\bolK_2-\bolK_1){\cdot} \bolr)\\
&\times\frac{\rho_{\text{B}1}^\dagger\rho_{\text{B}2}}{N}\sum_{\bolp}\chi^\dagger_{\bolK_1}(\bolp)\,
\chi_{\bolK_2}\! \left(\frac{\bolK_2-\bolK_1}{2}+\bolp\right)+\text{h.c.} \biggr\}  \ ,
\end{split}
\end{equation}
where we have kept the terms up to the second order 
in $\rho_{\text{B}1,\text{B}2}$. 
Now, $\VEV{S^z_l}$ oscillates while the transverse magnetization vanishes. 
This phase may emerge through a second order phase transition 
from the usual nematic phase with $\rho_{\text{B}2}=0$, 
by continuously introducing a finite pair-condensate $\rho_{\text{B}2}$ at $\bolK_2$, 
in a similar way as the superfluid-supersolid phase transition 
in the hard-core boson model, 
where 
a roton minimum softens\cite{Scaletter}. 
Obviously, $\rho_{\text{B}1}$ corresponds to the pairs at $\bolK_1=(\pi,\pi)$. 
Possible candidates of the $\rho_{\text{B}2}$ pair may be the one formed by 
magnons near the single-particle minima $\pm \bolQ_s$. 
There are three candidates of $\mathbf{K}_{2}$: 
$\bolK_{2a}=\mathbf{0}$ for $(\bolQ_s,-\bolQ_s)$ and  
$\bolK_{2b}=\pm2\bolQ_s$ for $\pm(\bolQ_s,\bolQ_s)$  
\footnote{If the gap of later pairs of $\pm\bolK_{2}=\pm2\bolQ_s$ closes, 
there are two possibility whether 
one of $\pm\bolK_2$ or both condense. 
Then, the emergent phase could be 
understood by the interaction between the pairs.}. 
A possible scenario is as follows.  
Immediately below the saturation field, 
the usual single-$\bolK$ nematic phase appears first, and, by further reducing 
the external field, the second-order phase transition takes place where 
bound states 
at $\bolK=\bolK_2$ start to 
condense. 
A quantitative estimate of the critical field remains to be an open problem.

{\it Summary-}
We have discussed the properties of the nematic phase 
slightly below the saturation field by microscopically calculating the interaction 
$\Gamma^{(2)}$ between the magnon bound states. 
First, we have derived the general expression of $\Gamma^{(2)}$ as 
eq.~(\ref{Ch2:effectiveIntBM})  
at the first order in the two magnon scattering process. 
Using $\Gamma^{(2)}$ thus obtained, we have investigated the stability, 
the asymptotic behavior of the magnetization curve, 
and the low-energy excitation in the nematic phase.  
We have applied this method to analyze the magnetic properties of $S=1/2$ frustrated spin system 
LiCuVO$_4$. We have found that the nematic phase is stable near  
the saturation field and exhibits a steep magnetization curve reminiscent of that in 1D. 
This steep behavior slightly below the {\it actual} 
saturation field is consistent
with the recent NMR measurement \cite{Takigawa_new}.
We have also discussed the possible second-order-phase transition 
from the usual nematic phase to a new spin-density-modulated phase with the transverse nematic order. 

The authors thank D. Jido, Y. Nishida, and M. Takigawa for stimulating discussions.  
One of the authors (K.T.) was supported by 
Grants-in-Aid for Scientific Research
No. (C) 24540402 from MEXT, Japan.



\begin{widetext}
\section{Supplementary Materials: Derivation of the s-wave scattering amplitude between bound magnons}

In this supplemental note, we detail the derivation of 
the scattering amplitude $\Gamma^{(2)}$ between two-bound magnons. 
First, we briefly review the method to obtain the 
energy and the wavefunction of a bound state from 
the two-body Green's function. 
Next, we derive $\Gamma^{(2)}$ from the scattering theory.

\subsection{Energy and wavefunction of a stable bound state}
\label{Ch2:BMenergywave}
Generally, if we write $\ket{n_a}$ as the eigenstate of the Hamiltonian, 
the Green's function of the operator $O_{1,2}$ is given by, 
\begin{equation}
iG(O_1O_2;E)=\int_0^\infty\VEV{O_1(t)O_2(0)}e^{iE t}dt=i\sum_a\frac{\bra{\Omega}O_1\ket{n_a} \bra{n_a}O_2\ket{\Omega}}{E-E_{n_a}+i0^+}\ ,
\label{Ch2:GOO}
\end{equation}
where $\ket{\Omega}$ is the ground state and $O_{1,2}(t)=e^{iHt}O_{1,2}e^{-iHt}$. 
In the above equation, a pole exists at $E=E_{n_a}$ which implies the existence of the stable state. 

To study the two-body scattering problem in our case, 
we assume
$O({\bolK},\bolp)=a_{\bolK/2+\bolp}a_{\bolK/2-\bolp}$, 
$O_1=O({\bolK},\bolp^\prime)$, 
$O_2=O^\dagger({\bolK},\bolp)$. 
Now, the bare time-ordered-two-body-Green's function without interactions reads
\begin{equation}
\begin{split}
iG^{(2)}_0(E,\bolK;\bolp,\bolp^\prime)=&\int_{-\infty}^{\infty}dt e^{iEt}
\VEV{T(O(t;\bolK,\bolp^\prime)O^\dagger(0;\bolK,\bolp))}_0\\
=&\int\frac{d\omega}{2\pi}iG_0(\omega,\bolK/2+\bolp)iG_0(E-\omega,\bolK/2-\bolp)(\delta_{\bolp,\bolp^\prime}+\delta_{\bolp,-\bolp^\prime})\\
=&\frac{i}{E-(\omega(\bolK/2+\bolp)+\omega(\bolK/2-\bolp)-2\mu)+i0^+}
\frac{(N\delta_{\bolp,\bolp^\prime}+N\delta_{\bolp,-\bolp^\prime})}{N}
\ ,
\end{split}
\end{equation}
where the one-particle Green's function $iG_0$ is given by
\begin{equation}
iG_0(\omega,\bolq)=\frac{i}{\omega-\omega(\bolq)+i0^+}\ ,
\end{equation}
and $\omega(\bolq)$ is given by eq.~(4) in the main paper. 
The branch cut exists for $E=\omega(\bolK/2+\bolp)+\omega(\bolK/2-\bolp)-2\mu$, which represents the continuum of two particles. 

In the interacting case, the two particle Green's function in the fully saturated phase is exactly given by
\begin{equation}
\begin{split}
&iG^{(2)}(E,\bolK;\bolp,\bolp^\prime)=iG^{(2)}_0(E,\bolK;\bolp,\bolp^{\prime})\\
&+\frac{1}{N}\sum_{\bolp^{\prime\prime},\bolp^{\prime\prime\prime}}\frac{1}{4}iG^{(2)}_0(E,\bolK;\bolp,\bolp^{\prime\prime}) (-i\Gamma(\Delta=-E-2\mu,\bolK;\bolp^{\prime\prime},\bolp^{\prime\prime\prime})) iG^{(2)}_0(E,\bolK;\bolp^{\prime\prime\prime},\bolp^\prime)\ ,
\end{split}
\label{G2full}
\end{equation}
where $\Gamma$ represents the ladder diagram discussed in the main paper eq.~(5). 
While the pole (branch cut) of the first term in the right-hand side 
provides the continuum, 
$\Gamma$ in the second term may introduce a pole even below 
the continuum, which implies the existence of a stable bound state. 
Hence, the energy of the bound state $E_B=-\Delta_B-2\mu$ is determined by a pole of $M$. In the later discussion we will neglect the first term in eq.~(\ref{G2full}), which describes a non-scattered amplitude and eventually vanishes for $\bolp\neq \pm\bolp^\prime$. For convenience, sometimes we will abbreviate the notation of this term as 
$G^{(2)}(E,\bolK;\bolp,\bolp^\prime)=(1/4)G^{(2)}_0(\bolp) \Gamma(\bolp,\bolp^\prime) G^{(2)}_0(\bolp^\prime)$.

This Green's function is also related to the wave function of the bound state.
Near the pole of the bound state, the nearly diverging term is dominant, 
and we neglect the other terms. 
Hence, if we assume $\chi_{\bolK}(\bolp)$ as the wavefunction of the bound state, we obtain for $E\sim E_B$:
\begin{equation}
\begin{split}
G^{(2)}(E,\bolK;\bolp,\bolp^\prime)& 
\approx\frac{\bra{\Omega}O(\bolK,\bolp^\prime)\ket{B} \bra{B}O^\dagger(\bolK,\bolp)\ket{\Omega}}{E-E_{B}(\bolK)+i0^+}=\frac{1}{N}\frac{\chi_{\bolK}(\bolp^\prime){\chi_{\bolK}}^\dagger(\bolp)}{E-E_B(\bolK)}\ ,
\end{split}
\label{Greenchi}
\end{equation}
where $\ket{B}$ is the eigenket of the bound state,
and $\chi_{\bolK}(\bolp)$ is the wavefunction given by 
$\sqrt{N}\bra{\Omega}O(\bolK,\bolp)\ket{B}$
since the vacuum $\ket{\Omega}$ is the exact ground state corresponding to the fully saturated phase even in the interacting case.
This wavefunction satisfies the normalization condition as $(1/N)\sum_\bolp |\chi_{\bolK}(\bolp)|^2=2$, where the summation over $p$ is taken in 
the Brillouin zone, since $\chi_{\bolK}(\bolp)=\chi_{\bolK}(-\bolp)$.

\subsection{Effective interaction between bound magnons}
\label{Ch2:effectiveInt}
In this subsection, we study the effective interaction up to the first order 
in the ladder diagram of two single magnons. 
For convenience, we introduce $d^\dagger_{\bolK}$ as the creation operator of the bound state:
\begin{equation}
d^\dagger_\bolK
=\frac{1}{2\sqrt{N}}\sum_{\bolp}\chi_{\bolK}(\bolp)a^\dagger_{\bolK/2+\bolp}
a^\dagger_{\bolK/2-\bolp}\ ,\ \ d^\dagger_{\bolK}\ket{\Omega}=\ket{B}\ .
\end{equation}
The Green's function of the bound state for $E\sim E_B$ is given by
\begin{equation}
\begin{split}
&\int dt e^{iEt}\VEV{T(d_\bolK(t) d_\bolK^\dagger(0)}
=\frac{1}{4N}\sum_{\bolp,\bolp^\prime} iG^{(2)}(E,\bolK;\bolp,\bolp^\prime)
\chi_{\bolK}(\bolp)\chi_{\bolK}^\dagger(\bolp^\prime)\\
&=\frac{i}{16N}\sum_{\bolp,\bolp^\prime}\chi_{\bolK}^\dagger(\bolp^\prime)G_0^{(2)}(\bolp)\Gamma G_0^{(2)}(\bolp^\prime)\chi_{\bolK}(\bolp) 
=\frac{i}{4N^2}\sum_{\bolp,\bolp^\prime}\frac{|\chi_{\bolK}(\bolp)|^2|\chi_{\bolK}(\bolp^\prime)|^2}{E-E_B(\bolK)}\\
&=\frac{i}{E-E_B(\bolK)+i0^+}\ ,
\label{Ch2:GreenDD}
\end{split}
\end{equation}
where we use eq. (\ref{Greenchi}).


Let us discuss the scattering process of two-incoming-bound states. 
According to the LSZ-Reduction formula (for review, see Sec.~VII in Ref.~\onlinecite{PeskinSchroeder}), 
we consider the Fourier transformation of the correlation function:
\begin{equation}
\begin{split}
&\int_{-\infty}^{\infty} dt_1 e^{i\omega t_1}
\int_{-\infty}^{\infty} dt_2 e^{i\omega t_2}
\int_{\infty}^{\infty} dt_3 e^{-i\omega t_3}
\int_{\infty}^{\infty} dt_4 e^{-i\omega t_4}
\bra{\Omega}T\{d_{\bolK_1}(t_1) d_{\bolK_2}(t_2) d^\dagger_{\bolK_3}(t_3) d^\dagger_{\bolK_4}(t_4)\}\ket{\Omega}\\
&\approx
\int_{T^+}^{\infty} dt_1 e^{i\omega t_1}
\int_{T^+}^{\infty} dt_2 e^{i\omega t_2}
\int_{-\infty}^{T^-} dt_3 e^{-i\omega t_3}
\int_{-\infty}^{T^-} dt_4 e^{-i\omega t_4}
\bra{\Omega}d_{\bolK_1}(t_1) d_{\bolK_2}(t_2) d^\dagger_{\bolK_3}(t_3) d^\dagger_{\bolK_4}(t_4)\ket{\Omega},
\end{split}
\end{equation}
where we assume $T^+>>T^-$ and $T^+$ ($T^-$) is in the `out' (`in') time region where 
the interaction between bound magnons is asymptotically zero 
since we consider the case that each bound magnon is well separated after and before the scattering process. 
Diagrammatically this process is represented as the left diagram in Fig.~\ref{Fig:boundInt}, which gives:
\begin{equation}
\frac{-i\Gamma^{(2)}}{N}\Pi_{l=1,2,3,4}(\frac{i}{\omega_{l}-E(\bolK_l)+i0^+})\big|_{\omega_{l}\rightarrow E(\bolK_l)} \ ,
\label{swaveL}
\end{equation}
where $\Gamma^{(2)}$ is the scattering amplitude between the two bound magnons, 
and $\omega_{l}\rightarrow E(\bolK_l)$, $\bolK_l\rightarrow \bolQ_B$ is assumed 
since 
we consider the scattering between the physically stable lowest energy 
bound states. 
In this case, the field strength is zero because the fully saturated 
ferromagnetic phase is the exact ground state and the self energy of the bound state is zero.
From another viewpoint, eq.~(\ref{swaveL}) is easily understood from the effective Hamiltonian 
eq.~(8) in the main paper.

\begin{figure}[ht]
\begin{center}
\includegraphics[scale=0.7]{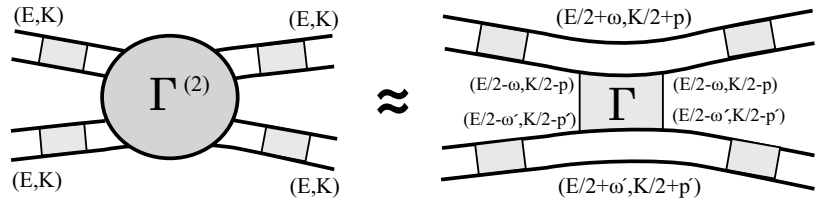}
\caption{Diagram of the two-body scattering of the bound magnons. 
The highlighted rectangles represent the two-body scattering amplitude 
$\Gamma$ of the single magnons. 
\label{Fig:boundInt}}
\end{center}
\end{figure}

To calculate $\Gamma^{(2)}$, we consider the first-order expansion 
in the $\Gamma$ (the scattering amplitude between the two single magnons) 
as shown in the right diagram in Fig.~\ref{Fig:boundInt}.  
The external lines are calculated as:
\begin{equation}
\begin{split}
&\frac{1}{N}\sum_{\bolp,\bolp^\prime}
\Bigl[\bigl\{\frac{1}{4\sqrt{N}}\sum_{\bolp^{\prime\prime}}(-i\Gamma(\bolp^{\prime\prime},\bolp))iG_0^{(2)}(\bolp^{\prime\prime})\chi_{\bolQ_\bolK}(\bolp^{\prime\prime})\bigr\}
\bigl\{\frac{1}{4\sqrt{N}}\sum_{\bolp^{\prime\prime}}(-i\Gamma(\bolp^{\prime\prime},\bolp^\prime))iG_0^{(2)}(\bolp^{\prime\prime})\chi_{\bolQ_\bolK}(\bolp^{\prime\prime})\bigr\}\\
&\ \ \ \ \ \ \ \times 4T(\bolQ_K;\bolp,\bolp^\prime)\\
&\times
\bigl\{\frac{1}{4\sqrt{N}}\sum_{\bolp^{\prime\prime}}\chi_{\bolQ_\bolK}^\dagger(\bolp^{\prime\prime})iG_0^{(2)}(\bolp^{\prime\prime})(-i\Gamma(\bolp,\bolp^{\prime\prime}))\bigr\}\bigl\{\frac{1}{4\sqrt{N}}\sum_{\bolp^{\prime\prime}}\chi_{\bolQ_\bolK}^\dagger(\bolp^{\prime\prime})iG_0^{(2)}(\bolp^{\prime\prime})(-i\Gamma(\bolp^\prime,\bolp^{\prime\prime}))\bigr\}\Bigr]\\
=&\frac{1}{N^3}\sum_{\bolp,\bolp^\prime}|\frac{1}{4}\sum_{\bolp^{\prime\prime}}(-iG^{(2)-1}_0(\bolp))(iG^{(2)}_0(\bolp))(-i\Gamma(\bolp,\bolp^{\prime\prime})iG_0^{(2)}(\bolp^{\prime\prime})\chi_{\bolQ_\bolK}(\bolp^{\prime\prime})|^2\\
&\times |\frac{1}{4}\sum_{\bolp^{\prime\prime}}(-iG^{(2)-1}_0(\bolp^\prime))(iG^{(2)}_0(\bolp^\prime))(-i\Gamma(\bolp^\prime,\bolp^{\prime\prime})iG_0^{(2)}(\bolp^{\prime\prime})\chi_{\bolQ_\bolK}(\bolp^{\prime\prime})|^2\times 4T(\bolQ_K;\bolp,\bolp^\prime)\\
=&\frac{1}{N^3}\sum_{\bolp,\bolp^\prime}
|2G^{(2)-1}_0(\bolp)\frac{\chi_{\bolQ_\bolK}^\dagger(\bolp)}{\omega_0-E({\bolQ_\bolK})}|^2\times|2G^{(2)-1}_0(\bolp^\prime)\frac{\chi_{\bolQ_\bolK}^\dagger(\bolp^\prime)}{\omega_0-E({\bolQ_\bolK})}|^2
\times T(\bolQ_K;\bolp,\bolp^\prime)\\
=&\frac{1}{N}(\frac{i}{\omega_0-E(\bolQ_\bolK)})^4\frac{1}{N^2}\sum_{\bolp,\bolp^\prime}(\frac{iG^{(2)}_0(\bolp)}{2})^{-2}(\frac{iG^{(2)}_0(\bolp^\prime)}{2})^{-2}|\chi_{\bolQ_\bolK}(\bolp)|^2|\chi_{\bolQ_\bolK}(\bolp^\prime)|^2
\times T(\bolQ_K;\bolp,\bolp^\prime)\ ,
\end{split}
\end{equation}
where $\omega_0\rightarrow E(\bolQ_\bolK)$ is taken, 
and $T(\bolQ_K;\bolp,\bolp^\prime)$ is the diagram describing the internal scattering. $4$ in front of $T$ counts the permutation of the legs.
The internal diagram $T$ is given by:
\begin{equation}
\begin{split}
T(&\bolQ_K;\bolp,\bolp^\prime)=\int\frac{d\omega d\omega^{\prime}}{(2\pi)^2}(-i\Gamma(\Delta=-(E(\bolQ_{\bolK})-\omega-\omega^{\prime})-2\mu,\bolQ_{\bolK}-\bolp-\bolp^\prime;\frac{\bolp^\prime-\bolp}{2},\frac{\bolp^\prime-\bolp}{2}))\\
&\times\frac{i}{\frac{E}{2}+\omega-(\omega(\frac{\bolQ_{\bolK}}{2}+\bolp)-\mu)+i0^+}\frac{i}{\frac{E}{2}+\omega^{\prime}-(\omega(\frac{\bolQ_{\bolK}}{2}+\bolp^\prime)-\mu)+i0^+}\\
&\times (\frac{i}{\frac{E}{2}-\omega-(\omega(\frac{\bolQ_{\bolK}}{2}-\bolp)-\mu)+i0^+})^2(\frac{i}{\frac{E}{2}-\omega^{\prime}-(\omega(\frac{\bolQ_{\bolK}}{2}-\bolp^\prime)-\mu)+i0^+})^2\\
=&(-i\Gamma(\Delta=-(2E(\bolQ_{\bolK})-\omega(\frac{\bolQ_{\bolK}}{2}+\bolp)-\omega(\frac{\bolQ_{\bolK}}{2}+\bolp^\prime))-2\mu,\bolQ_{\bolK}-\bolp-\bolp^\prime;\frac{\bolp^\prime-\bolp}{2},\frac{\bolp^\prime-\bolp}{2}))\\
\times &(\frac{i}{E-(\omega(\frac{\bolQ_{\bolK}}{2}+\bolp)+\omega(\frac{\bolQ_{\bolK}}{2}-\bolp)-2\mu)+i0^+})^2(\frac{i}{E-(\omega(\frac{\bolQ_{\bolK}}{2}+\bolp^\prime)+\omega(\frac{\bolQ_{\bolK}}{2}-\bolp^\prime)-2\mu)+i0^+})^2\\
=&(-i\Gamma(\Delta,\bolQ_{\bolK}-\bolp-\bolp^\prime;\frac{\bolp^\prime-\bolp}{2},\frac{\bolp^\prime-\bolp}{2}))(\frac{iG_0^{(2)}(\bolp)}{2})^2(\frac{iG_0^{(2)}(\bolp^\prime)}{2})^2\ .
\end{split}
\end{equation}
When we evaluate the residue integrals in $T$, we take the contour in the lower half-plane of $\omega$ to evade the pole of $\Gamma$ (see eq.(\ref{Ch2:GreenDD})).

In total, the effective interaction between the condensed bound state is given by 
\begin{equation}
\begin{split}
&\Gamma^{(2)}=\frac{1}{N^2}\sum_{\bolp,\bolp^\prime}|\chi_{\bolQ_\bolK}(\bolp)|^2|\chi_{\bolQ_\bolK}(\bolp^\prime)|^2\\
&\times \Gamma(2\Delta(\bolQ_{\bolK})+\omega(\frac{\bolQ_{\bolK}}{2}+\bolp)+\omega(\frac{\bolQ_{\bolK}}{2}+\bolp^\prime),\bolQ_{\bolK}-\bolp-\bolp^\prime;\frac{\bolp^\prime-\bolp}{2},\frac{\bolp^\prime-\bolp}{2})\ ,
\label{Ch2:effectiveIntBM}
\end{split}
\end{equation}
where we use $E(\bolQ_\bolK)=-\Delta(\bolQ_\bolK)-2\mu$.
We note that the integration should be carried out over the Brillouin zone 
if the bound state is normalized as $\frac{1}{N}\sum_\bolp |\chi_{\bolQ_\bolK}(\bolp)|^2=2$.


\end{widetext}
\end{document}